\documentclass[a4,12pt]{article}
\newcommand{\be}{\begin{eqnarray}}
\newcommand{\ee}{\end{eqnarray}}
\makeatletter
\newcommand{\fslash}[2][0mu]{%
    \mathchoice
     {\fsl@sh\displaystyle{#1}{#2}}%
     {\fsl@sh\textstyle{#1}{#2}}%
     {\fsl@sh\scriptstyle{#1}{#2}}%
      {\fsl@sh\scriptscriptstyle{#1}{#2}}}
    \newcommand{\fsl@sh}[3]{%
    \m@th\ooalign{$\hfil#1\mkern#2/\hfil$\crcr$#1#3$}}
\def\lsim{\raise0.3ex\hbox{$<$\kern-0.75em\raise-1.1ex\hbox{$\sim$}}}
\def\gsim{\raise0.3ex\hbox{$>$\kern-0.75em\raise-1.1ex\hbox{$\sim$}}}
\makeatother
\usepackage{epsfig}
\usepackage{graphics}

  \title{\bf Hadron Resonance Mass Spectrum  and  \\
Lattice QCD Thermodynamics\footnote{ ~Dedicated to Rolf Hagedorn}
   }
  \author{ \\ F. Karsch$^1$,
  K. Redlich$^{1,2}$, and A. Tawfik${^1}$
    \\ \\
    \small
 $^1$ Fakult\"at f\"ur Physik, Universit\"at Bielefeld,\\
\small Postfach 100 131, D-33501 Bielefeld, Germany\\
  \small     $^2$ Institute of Theoretical Physics University of Wroclaw,\\
\small   PL-50204 Wroclaw, Poland }

\begin{document}
\maketitle

\centerline{Abstract}
{\small
 We confront lattice QCD results on the transition from the hadronic phase
to the quark--gluon plasma with hadron resonance gas and
percolation models. We argue that for $T\leq T_c$ the equation of
state derived from Monte--Carlo simulations of (2+1) quark--flavor
QCD can be well described by a hadron resonance gas. We examine
the quark mass dependence of the hadron spectrum  on the lattice
and discuss its description in terms of the MIT bag model. This is
used to formulate a resonance gas model for arbitrary quark masses
which can be compared to lattice calculations. We finally apply
this model to analyze the quark mass dependence of the critical
temperature obtained in lattice calculations. We show that the
value of $T_c$ for different quark masses agrees with lines of constant
energy density in a hadron resonance gas. For large quark masses
a corresponding contribution from a glueball resonance gas is required.
 }

\newpage
\section{Introduction}

Long before lattice calculations provided first evidence
\cite{first} for critical behavior in strongly interacting matter
it has been noticed  \cite{Hagedorn} by Hagedorn  that ordinary
hadronic matter cannot persist as a hadronic resonance gas at
arbitrary high temperatures and densities.  This lead to the
concept of the Hagedorn limiting temperature.  With the
formulation of QCD it has been suggested \cite{Cabibbo} that a
phase transition to a new form of matter, the quark--gluon plasma,
will occur.

Two basic properties of hadrons were essential for developing the
concept of a natural end for the era of ordinary hot and dense
hadronic matter. In high energy experiments it had been observed
that strongly interacting particles produce a large number of new
resonances. Moreover, hadrons have been known to be extended
particles with a typical size of about 1~fm. As the average energy
per particle increases at high temperatures copious particle
production will take place in a hadron gas and a dense
equilibrated system will result from this. At high temperature,
extended hadrons thus would start to ``overlap''.  This led to the
expectation that some form of new physics has to occur under such
conditions. The expected critical behavior has been analyzed in
terms of various phenomenological models  which incorporate these
basic features (resonance production $\Rightarrow$ Hagedorn's
bootstrap model \cite{Hagedorn}; extended hadrons $\Rightarrow$
percolation models \cite{percolation}). In fact, many of the basic
properties of the dense matter created today in heavy ion
experiments can be understood quite well in terms of the
thermodynamics of a hadronic resonance gas \cite{redlich,redb}.

With the formulation of Quantum Chromodynamics (QCD) as a
theoretical framework for the strong interaction force among
elementary particles it became clear that this ``new physics''
indeed meant a phase transition to a new phase of strongly
interacting matter -- the quark-gluon plasma (QGP) \cite{Cabibbo}.
As QCD is an asymptotically free theory, the interaction vanishes
logarithmically with increasing temperature, it has been expected
that at least at very high temperatures the QGP would effectively
behave like an ideal gas of quarks and gluons. Today we have a lot
of information from numerical calculations within the framework of
lattice regularized QCD about the thermodynamics of hot and dense
matter which give support to these expectations. We know about the
transition temperature to the QGP and the temperature dependence
of basic bulk thermodynamic observables such as the energy density
and the pressure \cite{review}. In the coming years the increase
in numerical accuracy certainly will lead to modifications of the
quantitative details of these results. However, already today they
are sufficiently accurate to be confronted with theoretical and
phenomenological models that provide  a description of
thermodynamics of strongly interacting matter. Recently,  progress
has been made
 to develop and link an improved perturbation
theory of QCD with lattice data  on the equation of state in the
deconfined phase \cite{pqcd}. In this paper we analyze in how far
the critical behavior can be understood in terms of the
physical degrees of freedom of the confined phase, {\i i.e.} those
of a hadronic resonance gas, and the intuitive percolation
picture \cite{Satz}.

Quite distinct from the phenomenological approaches to the
QCD phase transition are attempts to understand the thermodynamics
of strongly interacting matter in terms of low energy effective
theories, {\it i.e.} chiral perturbation theory \cite{Leutwyler}
and effective chiral models \cite{chiralmodel,ch2}. The strength
of these approaches is that they incorporate the correct
symmetries of the QCD Lagrangian and thus have a chance to predict
the universal properties, e.g. the order of the phase transition,
in the chiral limit of QCD. They, however, generally ignore the
contributions of heavier resonances to the QCD thermodynamics
which might be crucial for the transition to the plasma phase at
non-vanishing values of the quark masses.

Lattice calculations provide detailed information on the quark
mass dependence of the transition to the QGP as well as to the
hadron spectrum at zero temperature. In particular, we know that
the transition temperature drops substantially when decreasing the
quark mass from infinity (pure SU(3) gauge theory) to values close
to the physical quark mass. This drop in the critical temperature
can be understood at least qualitatively in terms of the relevant
degrees of freedom in the low temperature phase. In the pure gauge
limit this phase consists of rather heavy glueballs ($m_G \gsim
1.5$~GeV \cite{glu,g20,g25}). Quite a large temperature thus is
needed to build up a sufficiently large density of glueballs,
which could lead to critical behavior (percolation \cite{Satz}).
In the chiral limit, on the other hand, the low critical
temperature can be addressed to the presence of light Goldstone
particles, the pions, which can build up a large (energy) density
already at rather low temperatures. Along with this decrease of
the critical temperature goes an increase in the critical energy
density expressed in units of the critical temperature,
$\epsilon_c /T_c^4$, by about an order of magnitude. This reflects
the importance of new degrees of freedom in the presence of light
quarks. However, at the same time the critical energy density in
physical units (GeV/fm$^3$) turns out to be almost quark mass
independent.

In this paper we want to focus on these results. We will discuss
in how far the quark mass dependence of the transition temperature
found in lattice calculations is consistent with phenomenological
models and what this tells us about the influence of the chiral
sector of QCD on the transition temperature. In Section 2 we will
briefly summarize the formulation of hadron thermodynamics in terms
of a hadronic resonance gas. In section 3 we discuss the quark
mass dependence of the hadron spectrum and give a phenomenological
parametrization motivated by the bag model. Predictions of these
phenomenological approaches for the equation of state and the
quark mass dependence of the transition temperature are then
compared with lattice results in Section 4. Finally we give
our conclusions in Section 5.

\section{Hadron resonance gas and the equation of state on the lattice}

The basic quantity required to verify thermodynamic properties
of QCD is the partition function\footnote{We restrict our discussion to
the case of vanishing chemical potential (vanishing net baryon number)
and charge neutral systems.} $Z(T,V)$.
The grand canonical partition function is obtained as
 \be
Z(T,V) = {\rm Tr}[e^{-\beta H }] \quad ,
\label{eq3}
 \ee
where  $H$ is the Hamiltonian of the system and $\beta =1/T$ is
the inverse temperature. The confined phase of QCD we model as a
non--interacting gas of resonances -- the hadron resonance gas
model. To do so we use as Hamiltonian the sum of kinetic energies
of relativistic Fermi and Bose particles of mass $m_i$.
The main motivation of using this Hamiltonian is that it contains
all relevant degrees of freedom of the confined, strongly
interacting matter and implicitly includes interactions that
result in resonance formation \cite{Hagedorn}. In addition this
model was shown to provide a quite satisfactory description of
particle production in heavy ion collisions
\cite{redlich,redb,raf}.

With the above assumption on the
dynamics the partition function can be
calculated exactly and expressed as a sum over one--particle
partition functions $Z^1_i$ of all hadrons and resonances,

\be
 \ln Z(T,V)=\sum_i \ln Z_i^1(T,V).
\label{eqq1} \ee For particles of mass $m_i$ and  spin--isospin
degeneracy factor $g_i$ the one--particle partition function
$Z^1_i$ is given by,

\be \ln Z^1_i(T,V)= {{Vg_i}\over {2\pi^2}} \int_0^\infty
dpp^2\eta\ln (1+\eta e^{-\beta E_i}), \label{eqq2} \ee where $
E_i=\sqrt {p^2+m_i^2}$ is the  particle energy and   $\eta= -1$
for bosons and $+1$ for fermions.

Due to the factorization of the partition function in
Eq.~\ref{eqq1} the energy density and the pressure
of the hadron resonance gas,
\be
 \epsilon=\sum_i\epsilon_i^1~~,~~P=\sum_i P_i^1, \label{eqq3}
 \ee
are also
expressed as sums over single particle contributions $\epsilon_i^1$
and $P_i^1$, respectively. These are given by

\begin{eqnarray}
{{\epsilon_i^1}\over {T^4}} &=& \frac{g_i}{2\pi^2}\;
\sum_{k=1}^{\infty} \;(-\eta)^{k+1}\;\frac{(\beta m_i)^3}{k}
\;\left[\frac{3\;K_2(k\beta m_i)}{k\beta m} +
  \;K_1(k\beta m_i)\right]\label{eqq4} \\
\Delta_i^1&\equiv&\frac{\epsilon_i^1-3P_i^1}{T^4} = \frac{g_i}{2
\pi^2}\; \sum_{k=1}^{\infty} \;(-\eta)^{k+1}\;\frac{(\beta
m_i)^3}{k} \; K_1(k\,\beta m_i) \label{eqq5}
\end{eqnarray}
where $K_1$ and $K_2$ are modified Bessel functions.

\begin{figure}[htb]
\vskip -.2cm
\begin{minipage}[t]{49mm}
\includegraphics[width=14.89pc,height=16.5pc,angle=-90]{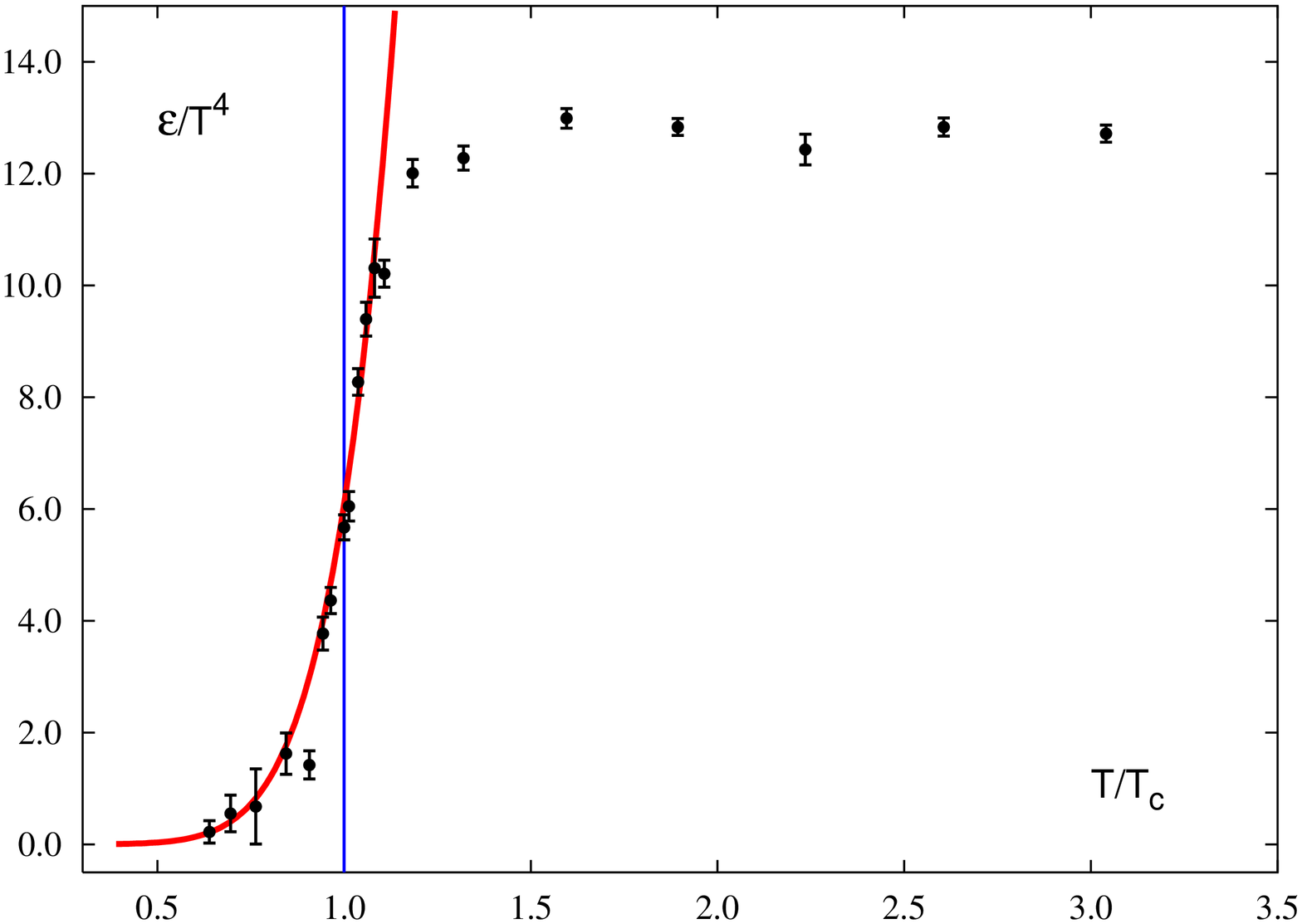}
\end{minipage}
\hskip 0.4cm
\begin{minipage}[t]{58mm}
\hskip 1.4cm
 \includegraphics[width=15.pc, height=16.3pc,angle=-90]{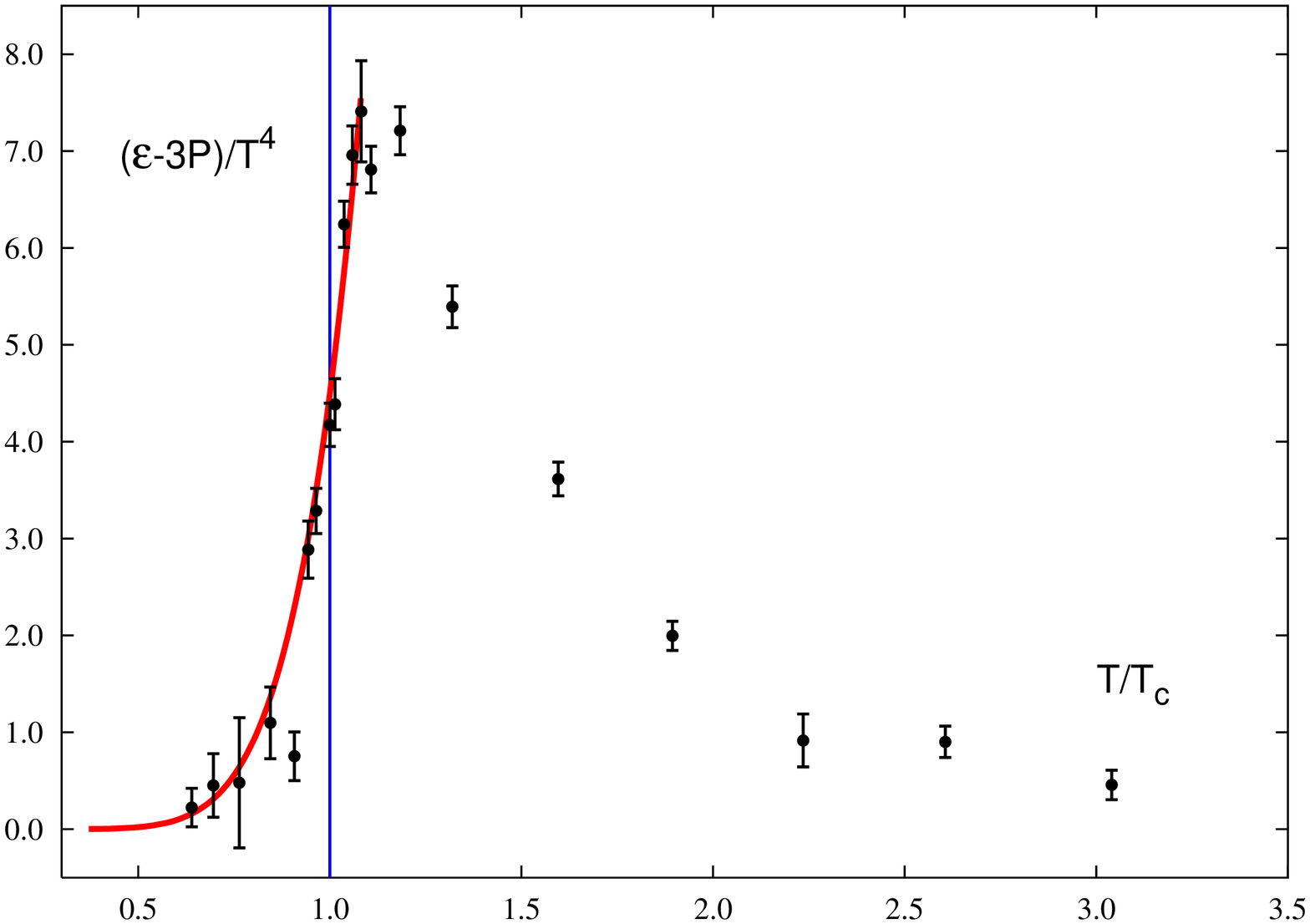}\\
\end{minipage}
\caption{ \label{fig1} The left--hand figure shows the energy
density $\epsilon$ in units of $T^4$ calculated on the lattice
with (2+1) quark flavors   as a function of the $T/T_c$ ratio. The
vertical lines indicate the position of the  critical temperature.
 The right--hand figure represents the corresponding
results for the interaction measure  $(\epsilon -3P)/T^4$. The
full--lines   are  the  results of the  hadron resonance gas model
that accounts for all mesonic and baryonic resonances.
 \hfill}
\end{figure}


Summing up in Eq.~(\ref{eqq3})  the contributions from
experimentally known hadronic states, constitutes
 the resonance gas model for
the thermodynamics of the low temperature phase of QCD. We take
into account all mesonic and baryonic resonances with masses up to
1.8 GeV and 2.0 GeV, respectively. This amounts to 1026
resonances. The energy density obtained in this way
%
starts rising rapidly at a temperature of about 160~MeV. In
Fig.~\ref{fig1} we show the temperature dependence of the energy
density $\epsilon$ and the interaction measure $\Delta$ for the
hadron resonance gas obtained from Eqs.~(\ref{eqq3}) to
(\ref{eqq5}). The model predictions are compared with Monte--Carlo
results obtained \cite{karsch1} on the lattice in (2+1) flavor
QCD. Although it should be noted that the lattice calculations
have not yet been performed with the correct quark mass spectrum
realized in nature the resonance gas model and the lattice data
agree quite well. This indicates that for $T\leq T_c$ hadronic
resonances are indeed the most important degrees of freedom in the
confined phase. The energy density in the resonance gas reaches a
value of 0.3~GeV/fm$^3$ at $T\simeq 155$~MeV and 1~GeV/fm$^3$
already at $T\simeq 180$~MeV. This is in good agreement with
lattice calculations, which find a critical energy density of
about 0.7~GeV/fm$^3$ at $T_c\simeq 170$~MeV \cite{karsch1}. For
comparison we note that a simple pion gas would only lead to an
energy density of about 0.1~GeV/fm$^3$ at this temperature. This
suggests that a more quantitative comparison between numerical
results obtained from lattice calculations and the resonance gas
model might indeed be meaningful.

\section{Hadron spectrum in heavy quark--mass limit }

In order to use the resonance gas model for further comparison
with lattice results we should take into account that
lattice calculations are generally performed with quark masses
heavier than those realized in nature. In fact, we should take
advantage of this by comparing lattice results obtained for
different quark masses with resonance gas model calculations based
on a modified, quark mass dependent, resonance spectrum.

Rather than converting the bare quark masses used in lattice
calculation into a renormalized mass it is much more convenient
to use directly the pion mass ($m_\pi \sim \sqrt{m_q}$), {\it
i.e.} the mass of the Goldstone particle, as a control parameter
for the quark mass dependence of the hadron spectrum.
%
%
For our thermodynamic considerations
we need, at present, not be concerned with the detailed structure of
the hadron spectrum in the light quark mass chiral limit. We
rather want to extract information on the gross features of the quark
mass dependence of a large set of resonances.
In order to study the quark mass dependence of hadron masses in
the intermediate  region between the chiral and heavy quark mass
limits we adopt here an approach that is based on the Hamiltonian
of the MIT bag--model \cite{bag1}. Although, in the original
formulation this Hamiltonian breaks explicitly chiral symmetry and
implies non--conservation of the axial--vector current it still
provides a satisfactory description of the hadron mass spectrum
that can be used for our thermodynamic considerations.

In the limit of a static, spherical cavity the energy of the bag of
radius R is given by

\be
E=E_V+E_0+E_K+E_M+E_E \quad.
\label{eq1}
\ee

The first two terms are due to quantum fluctuations and are
assumed to  depend only on the bag radius. The volume and the
zero--point energy terms have a generic form

\be E_V={4\over 3}\pi BR^3~~~,~~~ E_0=-{{Z_0}\over
R} \quad ,
\label{eq2}
\ee
where  $B$ is the bag constant and  $Z_0$ is a
phenomenological parameter attributed to the surface energy.

The quarks inside the bag contribute with their kinetic and rest
energy. Assuming $N$ quarks of mass $m_i$ the  quark  kinetic
energy is determined from

 \be
 E_K={1\over R}\sum_{i=1}^N[x_i^2+(m_iR)^2]^{1/2} \quad ,
 \label{eq4}
 \ee
where $x_i(m_i,R)$ enters the expression on the frequency $\omega
=[x^2+(mR)^2]^{1/2}/R$  of the lowest quark mode and is obtained
\cite{bag1} as the smallest positive root of the following
equation

 \be
 \tan(x_i)={{x_i}\over {1-m_iR-\sqrt{x_i^2+(m_iR)^2}}} \quad .
 \label{eq5}
 \ee

The last two terms in Eq.~(\ref{eq1}) represent the
color--magnetic and electric  interaction of quarks.  It is
described by the exchange of a single gluon between two quarks
inside the bag. The color electric energy was found in \cite{bag1}
to be numerical small and will be neglected in our further
discussion. The color magnetic exchange term is given by

\be E_M={8}k\alpha_c\sum_{i<j}{{M(m_iR,m_jR)}\over
R}(\vec{\sigma_i}\cdot\vec{\sigma_j}) \quad . \label{eq6}
 \ee
Here $\alpha_c$ is the strong coupling constant and $k=1$ for
baryons and 2 for mesons. For a given spin configuration of the
bag the scalar spin product in Eq.~(\ref{eq6}) can easily be
calculated. The function $M(x,y)$ depends on the quark modes
magnetic moment and is described in detail in \cite{bag1}. For
small $x<1$ it shows a linear dependence on the argument with
$M(0,0)=0.175$.

The dependence of the energy on the bag radius can be eliminated
by the condition that the quark and gluon field pressure balance
the external vacuum pressure. For a static spherical bag this
condition is equivalent to minimizing $E$ with respect to $R$. The
true radius $R_0$ of the bag thus is determined from the condition
$\partial E/\partial R=0$ and the hadron bag mass is then obtained
from Eq. (\ref{eq1})  with $R=R_0$.

To extract the physical mass spectrum from the MIT bag model one
still needs to fix the set of five parameters that determine  the
bag energy. Following the original fit to experimental data made
in \cite{bag1} we take: $B^{1/4}=0.145$ GeV, $Z_o=1.84$ and
$\alpha_c=0.55$. These parameters together with $m_u=m_d=0$ and
$m_s=0.279$ GeV provide a quite satisfactory description of hadron
masses belonging to the octet and decuplet of baryons and the
octet of vector mesons.
Of course, the model fails to describe the details of the chiral limit
and, in particular, it leads to
a too large value of the pion mass that with the above
set of parameters is $m_\pi=0.28$ GeV. Nonetheless, the accuracy
of the bag model will be sufficient for our purpose.
\begin{figure}[t]
\begin{center}
\vskip -2cm
\includegraphics[width=25.5pc, height=28.5pc,angle=180]{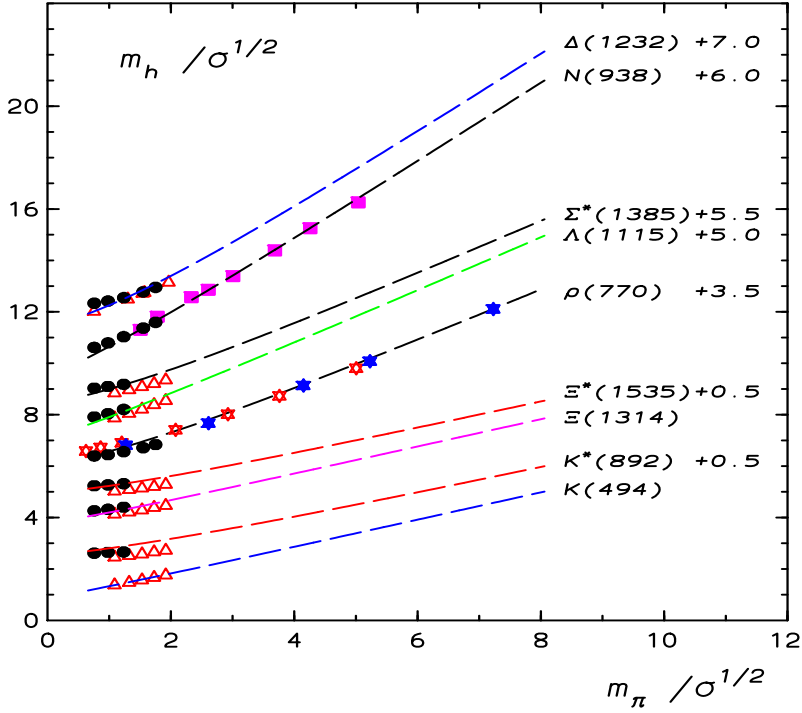}
\end{center}
\vskip -1.5truecm \caption{Dependence of different hadron masses
$m_h$ on the pion mass $m_\pi$.  Both $m_h$ and $m_\pi$ are
expressed in the units of the  string tension $\sqrt\sigma$.
Curves are the MIT bag model results (see text for details). The
filled circles represent the PC--PACS lattice results from
\cite{data1}. The filled diamonds are the $N_f=3$ whereas the
open--diamonds are $N_f=2$ flavor results from \cite{peikert1}.
The filled--boxes are quenched QCD results  \cite{data3}. All
other points are from reference \cite{data6}. Both the lattice
data and the bag model results are shifted in $m_h$--direction by
a constant factors indicated in the figure.
 \hfill}
\label{qm10}
\end{figure}

The MIT bag model provides an explicit dependence  of hadron
masses  on the constituent quark mass. This dependence is entirely
determined by the kinetic and magnetic energy of the quarks. To
compare bag model calculations with lattice calculations, which do
not provide values for constituent quark masses, it is best to
express the quark mass dependence in terms of the pion mass, which
is most sensitive to changes of the quark masses. In
Fig.~\ref{qm10} we show the resulting dependence of different
hadron masses on the pion mass with the bag parameters described
above but with varying $m_u$. The masses are expressed in units of
the square root of the string tension for which we use
$\sqrt{\sigma} = 420$~MeV.
 The model predictions are
compared with recent lattice data on hadron masses calculated for
different current quark masses \cite{peikert1,data1,data3,data6}.
The MIT bag model is seen in Fig.~\ref{qm10} to describe lattice
results quite well. This is particularly the case for  larger
quark masses such that $m_\pi >\sqrt\sigma$.
For $m_\pi <\sqrt\sigma$ the deviations of the
model from the lattice results are quite apparent. As mentioned
this is, of course, mainly due to the well known limitations of
the bag model when approaching the chiral limit.

For large quark masses the bag model description of hadron masses
reproduces the naive parton model picture and consequently all
hadron masses are almost linearly increasing with the pion mass as
seen in  Fig.~\ref{qm10} . This is to be expected as in this case
the energy of the bag is entirely determined by the quark rest
mass. As seen in  Fig.~\ref{qm10} the slope increases with the
number of non--strange constituent  quarks inside the bag.
Consequently, the slops of ($\Xi^*,\Xi$) and ($K,K^*$) or
($\Sigma^* ,\Lambda$) and $\rho$ coincide at large $m_\pi$.

In order to formulate a resonance gas model for arbitrary quark
masses we need to know the quark mass dependence of much more
resonances than the few hadronic states shown in Fig.~\ref{qm10}.
We thus looked for a phenomenological parametrization of the quark
mass dependence of resonances, expressed in terms of the pion
mass.
Fig.~\ref{qm10} suggests that already at intermediate values of the quark
mass, $m_\pi \gsim \sqrt{\sigma}$, this dependence is dominated
by the quark rest mass and does not depend much on the hadronic
quantum numbers. This suggests that a common parametrization of all
hadronic states, which is consistent with the naive parton model
picture for large quark masses and reproduces the experimental
values of hadronic states in the light quark mass limit is sufficient
for our thermodynamic considerations. To incorporate these features
we use the ansatz,


\begin{equation}
{{M(x)}\over{\sqrt\sigma}}\simeq N_ua_1x+{{m_0}\over
{1+a_2x+a_3x^2+a_4x^3+a_5x^4}} \quad ,
\label{par}
\end{equation}
which provides a good description of the MIT bag model result for
non--strange hadron masses calculated for different values of
$m_\pi$. Here $x\equiv {m_\pi /\sqrt\sigma}$, $m_0\equiv
{m_{hadron} /\sqrt\sigma}$, $N_u$ is the number of light quarks
inside the hadron ($N_u=2$ for mesons $N_u=3$ for baryons) and
$\sigma =(0.42$~GeV)$^2$ is the string tension.

The parameters appearing in  Eq.~(\ref{par}) were optimized
such that they reproduce the MIT bag model results for the
$m_\pi$--dependence of the $\rho$ vector meson mass and are
summarized in Table 1.
In the mass regime shown in Fig.~\ref{qm10} Eq.~(\ref{par})
reproduces the quark mass dependence of all  non--strange hadron
masses obtained from  the bag model within a relative error of
$\lsim$6$\%$.

\hskip 1cm
\begin{table}[h]{ \hskip 0.5cm
\begin{tabular}{|c|c|c|c||c|} \hline
   &       &   &    &   \\
 $a_1$    & ~ $ a_2$ ~  & $a_3 $  &  $a_4$ & $a_5$   \\
            &       &   &     &  \\ \hline
  &       &   &    &   \\
0.51$\pm$ 0.1 &   ~  $a_1N_u\over {m_0}$ ~~& 0.115$\pm$ 0.02
&-0.0223$\pm$ 0.008 &
 0.0028$\pm$ 0.0015 \\
 &       &   &    &   \\\hline
\end{tabular}}
\caption{Parameters entering the interpolation formula for non--strange
hadron masses given in Eq.~(12).}
\end{table}


In the following we will use Eq.~(\ref{par}) to formulate
a hadron resonance gas model with varying  quark masses. It will
then be compared with lattice calculation of QCD thermodynamics.
We will test, in particular, in how far this model can provide a
quantitative description of the transition temperature obtained on
the lattice for different quark masses.

\begin{figure}[t]
\begin{center}
\epsfig{file=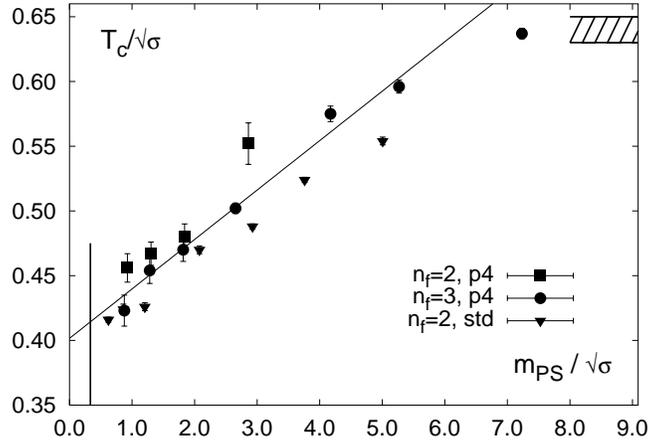,width=90mm}
\end{center}
\caption{The transition temperature in 2 (filled squares) and 3
(circles) flavor  QCD versus $m_{PS}/\sqrt{\sigma}$ using an
improved staggered fermion action (p4-action). Also shown are
results for 2-flavor QCD obtained with the standard staggered
fermion action (open squares). The dashed band indicates the
uncertainty on $T_c/\sqrt{\sigma}$ in the quenched limit. The
straight line is the fit given in Eq.~\ref{tcfit}.}
\label{fig:tc_pion}
\end{figure}

\section{Quark mass dependence of the QCD transition}

We want to confront here the resonance gas model developed in the
previous section with lattice results on the quark mass dependence
of the QCD transition temperature and use it to learn about the
critical conditions near deconfinement. Lattice calculations
suggest that this transition is a true phase transition only in
small quark mass intervals in the light and heavy quark mass
regime, respectively. In a broad intermediate regime, in which the
pion mass changes by more than an order of magnitude, the
transition is not related to any singular behavior of the QCD
partition function. Nonetheless, it still is well localized and is
characterized by rapid changes of thermodynamic quantities in a
narrow temperature interval. The transition temperature thus is
well defined and is determined in lattice calculations through the
location of maxima in response functions such as the chiral
susceptibility. A collection of transition temperatures obtained
in calculations with 2 and 3 quark flavors with degenerate masses
is shown in Fig.~\ref{fig:tc_pion}. The main feature of the
numerical results which we want to explore here is that the
transition temperature varies rather slowly with the quark mass.
In Ref.~\cite{karsch1} the almost linear behavior has been
described by the fit,
\begin{equation}
\biggl({T_c \over \sqrt{\sigma}}
\biggr)_{m_{PS}/\sqrt{\sigma}} =
0.4
+ 0.04(1)\; \biggl({m_{PS} \over \sqrt{\sigma}} \biggr)
\quad ,
\label{tcfit}
\end{equation}
which also is shown in Fig.~\ref{fig:tc_pion}. For pion masses
$m_{PS} \sim (6-7)\sqrt{\sigma} \simeq 2.5$~GeV the transition
temperature reaches the pure gauge value, $T_c/\sqrt{\sigma} \simeq
0.632 (2)$ \cite{qgp3}.

We note that all numerical results shown in Fig.~\ref{fig:tc_pion}
do correspond to quark mass values in the crossover regime. Also
the resonance gas model formulated in the previous section does
not lead to a true phase transition. We thus may ask what the
conditions in a hadron gas are that trigger the transition to the
plasma phase. Using the hadron gas with a quark mass dependent
hadron mass spectrum and including the same set of 1026 resonances
which have been included in other phenomenological calculations
\cite{redlich,redb} we have constructed resonance gas models for 2
and 3 flavor QCD, respectively. In the former case we eliminate
all states containing strange quarks whereas in the latter case we
assigned to meson states containing strange particles the
corresponding masses of non-strange particles, e.g. kaons have
been replaced by pions etc. With these resonance gas models we
have calculated the energy density at the transition temperature.
We use $T_c = 175\;(15)$~MeV for 2-flavor QCD and $T_c =
155\;(15)$~MeV for 3-flavor QCD, respectively. For the energy
densities at the transition point we then find
\begin{equation}
\biggl( {\epsilon \over T^4} \biggr)_{T=T_c} \simeq \cases{4.5\pm
1.5, & 2-flavor \cr 7.5 \pm 2, & 3-flavor \cr} \quad .
\label{epsilonres}
\end{equation}
This is in good agreement with the lattice result, $\epsilon
/T_c^4 = (6\pm 2)$ quoted in \cite{qm2002} as an average for the 2
and 3-flavor energy densities. In fact, as can be seen from Fig.5
in Ref.~\cite{qm2002} the difference in $\epsilon/T_c^4$ in the
lattice results is of similar magnitude as we found here from the
resonance gas model. The lattice results for 2 and 3-flavor QCD
thus suggest that the conditions at the transition point are well
described by a resonance gas. For comparison we also note that
in the 2-flavor case a pion gas does contribute only about 20\% to this
energy density\footnote{For massless pions we have $\epsilon /T^4
= (n_f^2-1) \pi^2/30 \simeq 1$.} and also a gas build up from the
20 lowest resonances would give rise only to about half  the
critical energy density, i.e. $\epsilon /T_c^4 \simeq 1.9$.

\begin{figure}[htb!]
  \begin{center}
    \includegraphics[width=6.7cm]{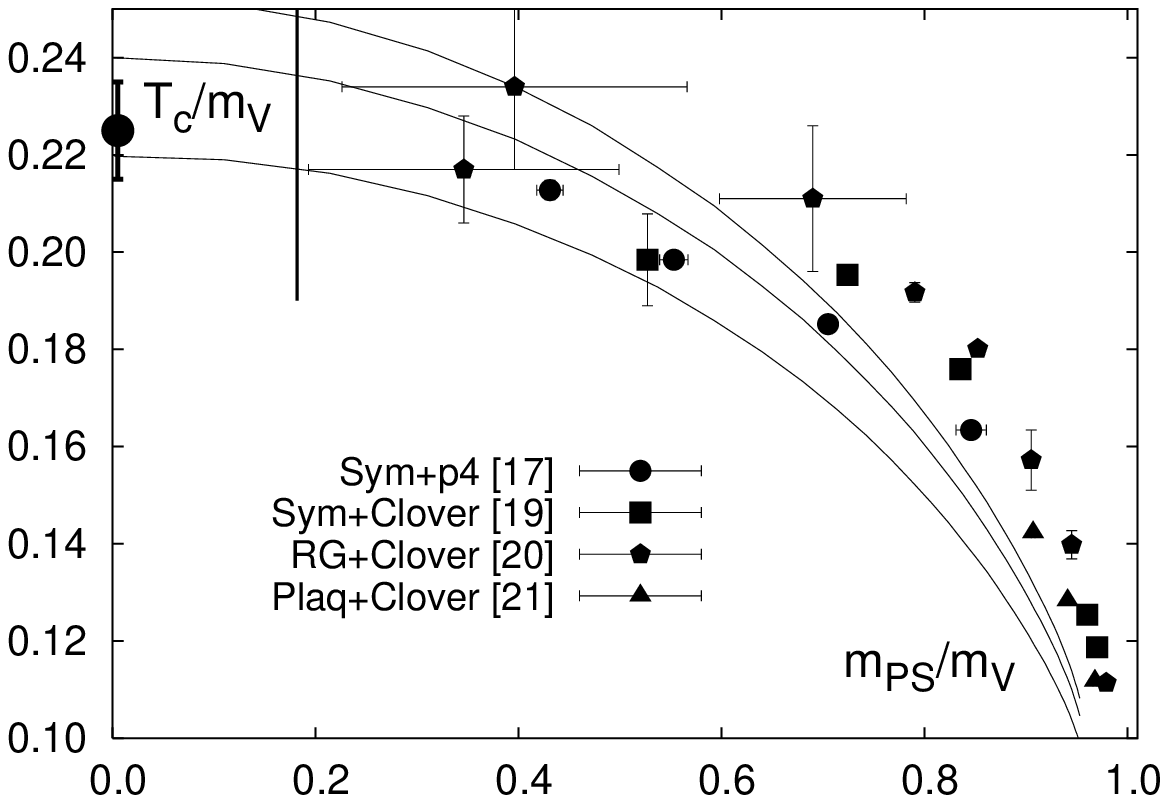}
    \includegraphics[width=6.7cm]{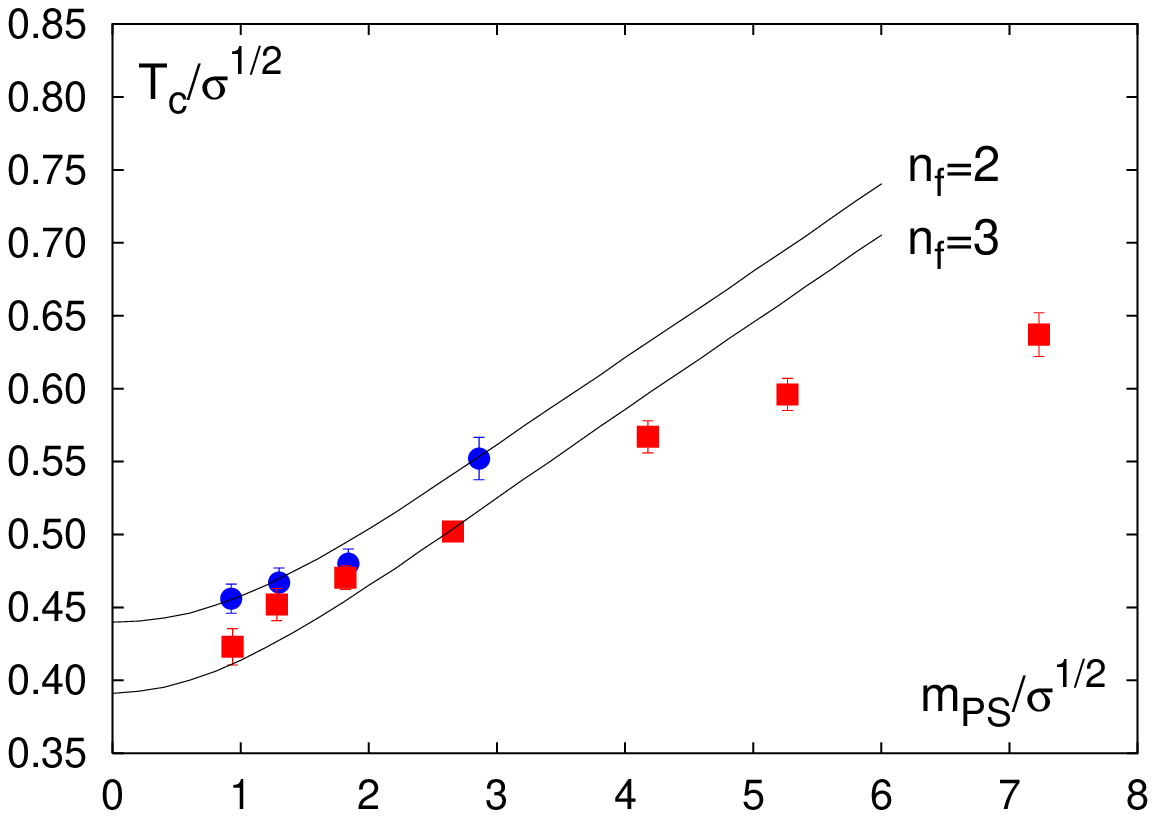}
    \caption{\small The  transition temperature vs. pion mass obtained
in lattice calculations and lines of constant energy density
calculated in a resonance gas model. The left hand figure shows a
comparison of constant energy density lines at 1.2 (upper),
0.8(middle) and 0.4(lower) GeV/fm$^3$ with lattice results for
2-flavor QCD obtained with improved staggered \protect\cite{karsch1}
as well as improved
Wilson \protect\cite{Bernard,Ali,Edwards} fermion formulations. $T_c$ as
well as $m_{PS}$ are expressed in terms of the corresponding
vector meson mass. The right hand figure shows results for 2 and 3
flavor QCD compared to lines of constant energy density of 0.8
GeV/fm$^3$. Here $T_c$ and $m_{PS}$ are expressed in units of
$\sqrt{\sigma}$. For a detailed description see text.}
    \label{fig:conditions}
  \end{center}
\end{figure}

Although the lattice results allow, at present, only to determine
the critical energy density within a factor (2-3) it is striking that
the transition occurs at similar values of the energy density in QCD
with light quarks as well as in the pure gauge theory, although the
transition temperature shifts by about 40\% and $\epsilon /T_c^4$
differs by an order of magnitude. It thus has been suggested that for
arbitrary quark masses the  transition occurs at roughly constant
energy density. Such an assumption is, in fact, supported by our
resonance gas model constructed in the previous section for arbitrary
values of the quark masses.
In Fig.~\ref{fig:conditions} we show lines of constant energy density
calculated in the resonance gas model and compare these to the
transition temperatures obtained in lattice calculations. As can be
seen the agreement is quite good up to masses, $m_{PS}\simeq
3\; \sqrt{\sigma}$ or $m_{PS}\simeq 1.2$~GeV.
The reason for the deviations at larger values of the quark mass,
of course, is due to the fact that we have neglected so far
completely the glueball sector in our considerations.

\begin{figure}[htb!]
  \begin{center}
    \includegraphics[width=8.0cm]{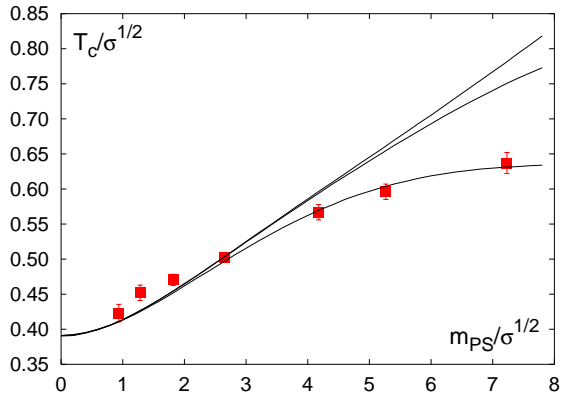}
    \caption{The transition temperature in 3-flavor QCD compared
to lines of constant energy density ($\epsilon = 0.8$~GeV/fm$^3$)
in a hadronic resonance gas (upper curve), a hadronic resonance
gas with 15 glueball states added (middle curve) and a hadronic
resonance gas with 15 glueball states with a  40\% reduced mass
(lower curve).}
    \label{fig:all}
  \end{center}
\end{figure}

When the lightest hadron mass becomes comparable to typical
glueball masses, also the glueball sector will start to contribute
a significant fraction to the energy density. Using the set of 15
different glueball states so far identified in lattice
calculations \cite{glu} we have calculated their contribution to
the total energy density. At $m_{PS}/\sqrt{\sigma} \simeq 6.5$
they contribute as much as the entire hadronic sector.
However, as can be seen in Fig.~\ref{fig:all} the contribution of
these 15 states
only leads to a small shift in the lines of constant energy density.
Similar to the hadronic resonance gas
for small quark masses where the 20 low-lying states only
contribute 50\% of the total energy density one has to expect that
also in the large
quark mass limit further glueball states, which have so far not
been identified, will contribute to the thermodynamics.
Further support for this comes from a calculation of the
energy density of the 15 known glueball states at the transition
temperature of the pure gauge theory, $T=0.63 \sqrt{\sigma}$.
For this we
obtain $\epsilon (T=0.63 \sqrt{\sigma}) \simeq =.06$ GeV/fm$^3$ or
equivalently $\epsilon/T_c^4 \simeq 0.1$, which is about 20\% of
the overall energy density at $T_c$.
The contribution of the 15 glueball states thus does not seem to
be sufficient. In fact, the transition temperature in $d$-dimensional
$SU(N_c)$ gauge theories is well understood in terms of the
critical temperature of string models,
$T_c/\sqrt{\sigma}=\sqrt{3/\pi(d-2)}$, which also is due to an
exponentially rising ''mass'' spectrum for string excitations
\cite{Pisarski}.


It is conceivable that extending the glueball mass spectrum to all
higher excited states will improve the results shown in
Fig.~\ref{fig:all}. On the other hand one also should stress that
the glueball states used in our calculations were obtained in
quenched QCD and at zero temperature. There are indications from
lattice calculations that glueball masses could be modified
substantially in the presence of dynamical quarks \cite{g25} as
well as at finite temperature \cite{g20}. The analysis of glueball
states at high temperature \cite{g20} suggests that their masses
can drop by $\sim (20-40)\%$. As all glueballs are heavy on
the temperature scale of interest, shifts in their masses influence
the thermodynamics much more strongly than in the
light quark mass regime where the lowest state has already a mass
which is of the order of the transition temperature. In fact, we
find that taking into account a possible decrease of the glueball
masses close to $T_c$ seems to be more important than adding further
heavy states to the spectrum. We thus have
included a possible reduction of glueball masses in the equation
of state. The resulting $T_c$ with this modification is also shown
in Fig.~\ref{fig:all}. Decreasing the glueball masses, increases
the thermal phase space available for particles, thus consequently
the temperature required to get $\epsilon =0.8$ GeV/fm$^3$ is
decreasing. As can be seen in Fig.~\ref{fig:all} a reduction of
glueball masses by 40$\%$ is sufficient to reproduce lattice
results in the whole $m_\pi$ range. However, to make this comparison
more precise it clearly is important to get a more detailed
understanding of the glueball sector in the future.

\section{Conclusions}

In this paper we have analyzed lattice results on QCD
thermodynamics using a phenomenological hadron resonance gas
model. We have shown that close to the chiral limit and for $T\leq
T_c$ the equation of state derived on the lattice is
quantitatively well described by the resonance gas.

The  hadron resonance gas partition function is also shown to be
suitable to describe lattice results for finite quark masses and
varying number of flavors. One needs, however, to implement the
quark mass dependence of the hadron  spectrum and for large values of
the quark mass the glueball degrees of freedom have to be taken into account 
as they start playing an
important role. We have shown that, away from the chiral limit region, 
the quark mass dependence arising from MIT bag model agrees quite well with
the hadron mass spectrum calculated on the lattice. We find that the
transition temperatures obtained  in lattice calculations at
different values of the quark mass are well described by lines of
constant energy density in a resonance gas model. For moderate
values of the quark masses the predictions of the hadron resonance
model coincide with lattice calculations. However, for heavy quark
masses this agreement could be only achieved   by including
additional heavy glueball states or allowing for a reduction  of
glueball masses close to the transition temperature by about
$40\%$.

Our results can be considered as  an indication that
thermodynamics in the vicinity of deconfinement is indeed driven
by the higher excited hadronic states. This finding can give
additional support for previous  phenomenological applications of
the resonance gas partition function in the description of particle
production in heavy ion collisions. Our discussion of the critical
temperature and its quark mass dependence also indicates that
deconfinement in QCD to large extend is density driven. It would
be interesting to see to what extent the lines of constant energy
density of the generalized hadron resonance gas can be related to
correspondingly generalized percolation models.

\section*{Acknowledgments}
\addcontentsline{toc}{section}{Acknowledgements}

We  acknowledge  stimulating discussions with  R. Baier, P.
Braun-Munzinger, E. Laermann, D. Miller and H. Satz. K.R. also
acknowledges the support of the Alexander von Humboldt Foundation
(AvH). This work has partly been supported by the DFG under grant
FOR 339/2-1.


\end{document}